\documentclass[twocolumn]{aastex63}
\usepackage{multirow}

\graphicspath{{./}{figures/}}

\begin{document}
\title{Detection of a $\sim$ 0.1\,c radio knot in M81* associated with a moderate X-ray flare}

\author{Xuezheng Wang}
\affiliation{Shanghai Astronomical Observatory, Chinese Academy of Sciences, 80 Nandan Road, Shanghai 200030, People's Republic of China; \href{jiangwu@shao.ac.cn}{jiangwu@shao.ac.cn}} 
\affiliation{ShanghaiTech University, 393 Middle Huaxia Road, Pudong, Shanghai, 201210, People's Republic of China}
\affiliation{University of Chinese Academy of Sciences, No.19(A) Yuquan Road, Shijingshan District, Beijing,100049, People's Republic of China}
\author[0000-0001-7369-3539]{Wu Jiang}
\affiliation{Shanghai Astronomical Observatory, Chinese Academy of Sciences, 80 Nandan Road, Shanghai 200030, People's Republic of China; \href{jiangwu@shao.ac.cn}{jiangwu@shao.ac.cn}}
\affiliation{Key Laboratory of Radio Astronomy and Technology, Chinese Academy of Sciences, A20 Datun Road, Chaoyang District, Beijing, 100101, People's Republic of China}

\author[0000-0003-3540-8746]{Zhiqiang Shen}
\affiliation{Shanghai Astronomical Observatory, Chinese Academy of Sciences, 80 Nandan Road, Shanghai 200030, People's Republic of China; \href{jiangwu@shao.ac.cn}{jiangwu@shao.ac.cn}}
\affiliation{Key Laboratory of Radio Astronomy and Technology, Chinese Academy of Sciences, A20 Datun Road, Chaoyang District, Beijing, 100101, People's Republic of China}
\author[0000-0002-5385-9586]{Zhen Yan}
\affiliation{Shanghai Astronomical Observatory, Chinese Academy of Sciences, 80 Nandan Road, Shanghai 200030, People's Republic of China; \href{jiangwu@shao.ac.cn}{jiangwu@shao.ac.cn}}
\author[0000-0002-7329-9344]{Ya-Ping Li}
\affiliation{Shanghai Astronomical Observatory, Chinese Academy of Sciences, 80 Nandan Road, Shanghai 200030, People's Republic of China; \href{jiangwu@shao.ac.cn}{jiangwu@shao.ac.cn}}
%
\author[0000-0003-3708-9611]{Ivan Mart\'i-Vidal}
\affiliation{Dpt. Astronomia i Astrof\'isica, Univ. Val\`encia,
C/Dr. Moliner 50, Agust\'in Escardino, E-46100 Burjassot, Spain}
\affiliation{Observatori Astron\`omic, Univ. Val\`encia,
C/Cat. Agust\'in Escardino 2,
E-16980 Paterna, Spain}
\author[0000-0003-2492-1966]{Roman Gold}
\affiliation{CP3-Origins, University of Southern Denmark, Campusvej 55, DK-5230 Odense M, Denmark}


\begin{abstract}


Through very long baseline interferometry observations of one of the closest low-luminosity active galactic nuclei M81* at multifrequencies of 8.8, 22 and 44\,GHz, a bright discrete knot with an unusual low apparent speed $\sim$0.1\,c was detected. Combining with the contemporary monitoring of X-rays data at 2-10\,keV, it indicates that a moderate X-ray flare happened when the knot launched from the core region. Three possible origins of the knot are proposed to explain our observational results. They are an episodic jet ejection, a low-speed shock wave, and a possible secondary black hole in a binary system, respectively. Future intensive multiwavelength monitoring can help to understand the discrete knot as well as the central black hole better.

\end{abstract}

\keywords{\href{http://astrothesaurus.org/uat/1390}{ Relativistic jets (1390)}; \href{http://astrothesaurus.org/uat/2033}{Low-luminosity active galactic nuclei(2033)};        \href{http://astrothesaurus.org/uat/1769}{Very long baseline interferometry(1769)}}


\section{Introduction} \label{sec:intro}

The nearby galaxy M81, located at a distance of 3.96$\pm$0.34 Mpc \citep{Bartel}, hosts a low-luminosity active galactic nucleus (LLAGN), M81*. M81* is the second-closest AGN, whose compact radio core has been detected at multiple frequencies up to 88\,GHz with very long baseline interferometry (VLBI) \citep{Jiang}. M81 was identified as a Seyfert\,1.5 galaxy, the subclass of Seyfert\,1 \citep{Ho1997}. This is in agreement with its face-on galactic disk in a ultraviolet observation \citep{Wang}, suggesting that the jet is in a small viewing angle. An angle of $\sim 14 \pm 2^{\circ}$ between the normal to the disk and the line of sight was estimated from the two-dimensional velocity field observed with HST \citep{devereux}. The bolometric luminosity was estimated to be $\sim3\times10^{41.5}$ erg/s from the spectral energy distribution, implying a low accretion rate of $1.7\times 10^{-5}\dot{M}_{\rm Edd}$ \citep{Ho}, where $\dot{M}_{\rm Edd}$ is the Eddington accretion rate. The jet is at a similar activity with a power of $10^{41.8}$ erg/s \citep{Quataert}. The core-shift effect generally attributed to synchrotron self-absorption can reflect the magnetic field and particle density of a jet \citep{Lobanov}. However, most of the time, M81* does not have a suitable nearby calibrator or an optically thin region as an anchor.  \citet{mart} take advantage of an ingenious phase-referencing target, SN\,1993J, to measure the core shifts at 1.7, 2.3, 5.0 and 8.4\,GHz. They estimated the magnetic field strength of the centimeter-wavelength VLBI cores with the core shift measurements. Typically, radio emissions from the core-jet structure \citep{Bietenholz} were observed to be steady, but occasional flares were also detected \citep{mart}. The radiation of X-rays had similar light curve \citep{Tomar}, where the flare flux at 2-10\,kev only double or triple as high as the quiescent stage. \citet{king} reported fast variability and a flare delay of more than 12 days between radio and X-ray in 2011. The time delay indicated a magnetic field strength at several Gauss, which was inconsistent with that estimated from the core shift. Additionally, during the flare, a rare outward component for such a low luminosity AGN was found with a moderate apparent speed of 0.51$\pm$0.17\,c. The periodic change in the position angle of the core provided strong evidence for jet precession \citep{mart,von}. This precession can generate variability in both the location of the core and the viewing angle.
                                                 
The connection between high-energy emissions and radio flare or knot ejection was investigated in some AGNs. For instance, \citet{Jorstad} discovered a strong correlation between $\gamma$-ray outbursts and optical, far-infrared, and radio emissions, occurring when the knot passed through the millimeter-wavelength VLBI core in the Blazar 3C\,454.3. A recent study showed a close correlation between nuclear 100\,GHz and intrinsic X-ray emissions in nearby radio-quiet AGNs, indicating that both millimeter-wavelength and X-ray emissions originate from the region known as the X-ray corona \citep{Ricci}. \citet{Rani} reported that high-energy flares were accompanied by the emergence of new VLBI components, with the timing of these events contingent upon the location of the high-energy emission region in 3C\,279.  

In this paper, we present analyses of new VLBI observations of M81* based a new discrete knot. The observational data and data reduction are described in Section \ref{sec:obs}; In Section \ref{sec:results}, we present the results of the core shift, the proper motion of the knot, and the polarization of the jet. We discuss the nature of the discrete knot in Section \ref{sec:discussion} and summarize the main conclusions in Section \ref{sec:summary}. The spectral index $\alpha$ is defined as $S_{\nu} \propto \nu^{+\alpha}$, where $S_{\nu}$ is the flux density measured at frequency $\nu$.  


\section{Observations and data reduction } \label{sec:obs}

We observed M81* with VLBA over three epochs from April 22 to June 10, 2018. Two of them were observed at frequencies 8.8, 22 and 44\,GHz, while the other was observed at 44\,GHz and 88\,GHz (we only include the 44\,GHz data in this paper). We used all ten VLBA antennas for the observations. Data were recorded at a recording rate of 2\,Gbps with dual-polarizations. The total bandwidths were 512\,MHz splitting into 16 intermediate frequency (IF) bands (each IF had a bandwidth of 32\,MHz). 
Bright sources OJ\,287 and 0954+658 were used as calibrators in all epochs. To monitor the proper motion of knot at the earlier stage, we also reanalyzed the data recorded on December 18, 2017 at 22\,GHz and February 19, 2018 at 15\,GHz from the public VLBA archive system. 

\begin{deluxetable*}{ccccccc}
\tablenum{1}
\tablecaption{Knot parameters \label{tab:knot}}
\tablewidth{0pt}
\tablehead{
\colhead{Epoch }& \colhead{Frequency }  & \colhead{Distance to the core } & \colhead{P.A.} & \colhead{Knot Flux } & \colhead{$T_b$ } &\colhead{Core Flux} \\
\colhead{(yyyy/mm/dd)}& \colhead{(GHz)}  & \colhead{(mas)} & \colhead{(degree)} & \colhead{(mJy)} & \colhead{(K)} &\colhead{(mJy)}
}
\decimalcolnumbers
\startdata
2017/12/18 &22&0.39&82.4&12.4&$2.1\times 10^9$ &84.0\\ 
\hline
2018/2/19  & 15.5& 0.68& 82.0& 15.9 & $3.2\times 10^8$ &84.4\\ 
\hline
2018/04/22  & 44 & 1.16 & 77.3 & 5.0& $1.3\times 10^8$ &102.0\\
\hline
\multirow{3}{*}{2018/05/14}&  8.8  & 0.92& 75.0 &  54.4 & $7.1\times 10^9$&127.0  \\
								  & 22  & 1.04 &75.8&  14.2& $2.6\times 10^8$ &73.0 \\ 
									& 44 & 1.29 &84.3&  5.8 & $7.6\times 10^7$ &96.1\\
\hline
\multirow{3}{*}{2018/06/10}&  8.8  & 1.14& 77.7 & 39.7 & $5.7\times 10^9$&95.6  \\
								  & 22  & 1.22 &78.3&  8.3& $1.3\times 10^8$ &79.0 \\ 
									& 44 & 1.31 &79.9&  3.0 & $3.9\times 10^7$ &66.3\\    
\enddata
\tablecomments{Columns (1)$\sim$(7): observation epoch (date); observation frequency; knot distance to the core; position angle of the knot; flux of the knot ; brightness temperature of the knot and flux of the core.}
\end{deluxetable*} 

The VLBA data were correlated at the National Radio Astronomy Observatory (NRAO) VLBA correlator of the Array Operations Center at Socorro (New Mexico). The data were calibrated using standard algorithms, as implemented in the NRAO Astronomical Image Processing System (AIPS) \citep{Greisen}. We then further refined the calibration using hybrid imaging \citep[e.g.][]{Cornwell}, as implemented in the program DIFMAP \citep{Shepherd}, which consists of iterating CLEAN deconvolution and visibility self-calibration until the chi-square converges to 1. Both the phase and amplitude were self-calibrated. MODELFIT function in DIFMAP was used to estimate the flux and position of the discrete knot on the basis of the calibrated visibilities. Table \ref{tab:knot} shows the parameters of the discrete knot fitted with a circular Gaussian component.

The high SNR ($\sim$250) of the discrete knot at 8.8\,GHz on June 10, 2018 offers a unique opportunity to study the linear polarization of the jet. Instrumental polarization was calibrated with the LPCAL task in AIPS. The linearly polarized flux $P= (U+Q)^{1/2}$ was obtained from the total U and Q fluxes in the polarization images. The right-left phase difference was calibrated by referring to the calibrator OJ\,287, whose polarization position angles were $128^{\circ}$ on May 31, 2018 from 15\,GHz MOJAVE program \footnote[1]{\href{http://cv.nrao.edu/MOJAVE}{http://cv.nrao.edu/MOJAVE}} and $137^{\circ}$ on June 16, 2018 from 43\,GHz VLBA-BU Blazar Monitoring Program \footnote[2]{\href{http://bu.edu/blazars}{http://bu.edu/blazars}}. We found that the EVPAs in these two epochs were consistent, suggesting that the observation frequency and date within ten days did not significantly affect the polarization angle. Since the observation frequency of 15\,GHz was closer to our data, we used the angle of $128^{\circ}$ for EVPA calibration. The polarization angle was calculated with rotated U and Q: ${\rm EVPA}= \frac{1}{2}\tan^{-1}(\frac{U}{Q})$.

We analyzed the archival observations of the X-ray Telescope (XRT) on board the Neil Gehrels {\em Swift} Observatory. The data reduction and spectral analysis are the same as Jiang et al. (2023, in preparation). We then obtained the X-ray flux in the 2--10 keV covering from 2016 to 2020 (see Table \ref{tab:xray}). The X-ray light curve in band 2-10\,keV observed by $Chandra$ \citep{Niu} from 2016-2018 are added to show the flux level at the quiescent stage.

\begin{deluxetable*}{cccc}
\tablenum{2}
\tablecaption{Swift X-ray observation of M81* in this study \label{tab:xray}}
\tablewidth{0pt}
\tablehead{
\colhead{OBSID}&\colhead{Start Date (yyyy/mm/dd)} & \colhead{MJD} & \colhead{Flux ($\rm 10^{-11}erg\, cm^{-2}\, s^{-1}$)}
}
\decimalcolnumbers
\startdata
00084311006&2016/10/19&57680.8& $1.70\pm0.17$ \\
00032519042&2017/06/10& 57914.4& $2.03\pm0.28$  \\
00032519043&2017/06/10 &57914.9&  $2.04\pm0.25$ \\ 
00032519044&2017/06/12&57916.4& $2.18\pm0.24$ \\
00084311008&2017/07/16&57950.7& $2.01\pm0.34$  \\
00084311012&2017/10/19 &58045.5&  $2.68\pm0.38$ \\ 
00084311024&2018/07/12&58311.0& $2.09\pm0.24$\\    
00095730018&2020/11/17&59170.1& $1.84\pm0.35$\\
00089156002&2020/11/19&59172.4&$1.35\pm0.12$\\ 
\enddata
\tablecomments{Columns (1)$\sim$(4): observation ID of $SWIFT$; observation date; observation date in MJD and the X-ray flux at 2-10\,kev band}
\end{deluxetable*} 
\section{Results} \label{sec:results}
\subsection{Core Shift} \label{subsec:core shift}

\begin{figure*}
	\centering
	\includegraphics[scale=0.7]{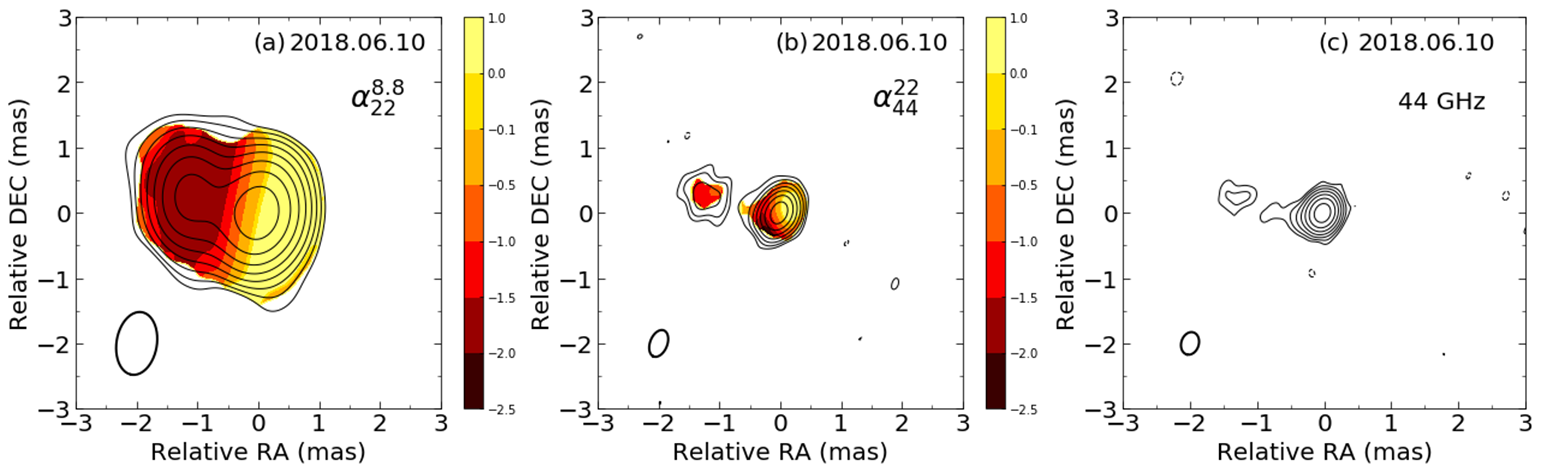}
	\caption{ VLBI images of M81* at 8.8, 22 and 44\,GHz in June 10, 2018.  (a) and (b) show the spectral index images with uniform weighting, while (c) shows the image with natural weighting. The contours are plotted at increasing powers of 2 from 0.37, 0.67, 0.61\,mJy for panels (a), (b), and (c), respectively. The beam is plotted in the lower left corner. 
 \label{fig:sp}  }
\end{figure*}

In all epochs, a discrete and compact knot is distinctly detected approximately 1\,mas downstream of the core. The knot is moving outward and fading, whose parameters are listed in Table \ref{tab:knot}. Compared to the knot observed in 2011 \citep{king}, our data present a knot that is visually more distinct from the core region. This indicates that such a knot is rare but repeatable observable (maybe periodic) case in M81*. The position of the knot is accurate in model-fitting due to its high SNR and discrete structure. By aligning the center position of the knot, we obtain the spectral index images of the jet as shown in Figure \ref{fig:sp}. Here we present the data of 10 June 2018, which not only had higher SNRs, but also showed consistency in the position angles of the knot across all frequencies. We check the spectral index images in another multi-frequency epoch, which show the expected behavior of a flat and steep (i.e., $\alpha \sim 0$ and $\alpha < 0$) spectrum for the core and knot, respectively.
As a region of synchrotron radiation, the position of the optically-thin knot is considered not to shift with frequencies for a given epoch, while the position of the core is frequency dependent due to synchrotron self-absorption \citep{Lobanov}. This knot offers a unique opportunity to study the core shift in M81*, which typically lacks suitable phase-referencing sources or a bright optically thin region.

\begin{figure}
	\centering
	\includegraphics[scale=0.5]{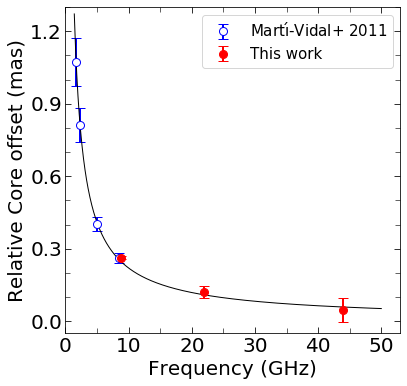}
	\caption{Relative location of the compact core as a function of observing frequency. The fitted model is shown in black. 
 \label{fig:core_shift}  }
\end{figure}

The core shift has been studied in \citet{mart}, whose highest frequency is 8.4\,GHz, almost coinciding with the lowest frequency of our data. The core shift between two frequencies $\nu_1, \nu_2\,(\nu_1 > \nu_2)$ was given as $\Delta r = \Omega (\frac{\nu_1^{1/k_r}-\nu_2^{1/k_r}}{\nu_1^{1/k_r} {\nu_2}^{1/k_r}})$ \citep{Lobanov}, $\Omega$ is the core shift measurement normalized to any pair of observing frequencies in units of pc·GHz (or mas·GHz); $k_r$ reflects the power laws of decreasing magnetic field and particle density. It is also the difference of the radial distance of the knot to the core: $\Delta r = d(\nu_1) - d(\nu_2)$ , where $d(\nu)$ is listed in column (3) in Table \ref{tab:knot}. Based on the core-shift result in \citet{mart}, we obtain new $r_{\rm core}$ values to fit the $r_{\rm core}-\nu$ relation. 
The separation from the central engine to the core at frequency $\nu$ is expressed as: $r_{\rm core}(\nu)=\Omega\nu^{-1/k_{\rm r}}$. Figure \ref{fig:core_shift} shows $r_{\rm core}(\nu)$ from 1.7 to 44\,GHz and the new fitted curve with $\Omega=1.72\,\rm mas\cdot GHz$, ${k_{\rm r}}=1.12$. The central engine is located at $r_{\rm core}(44)=0.06$\,mas upstream of the radio core of 44\,GHz. If assuming $k_{\rm r} =1$, $\Omega$ becomes about 1.75\,$\rm mas\cdot GHz$ (0.034\,$\rm pc\cdot GHz$), aligning well with the parameters estimated by \citet{mart}. We calculate the magnetic field at 1\,pc ($\rm B_1$) following the equation (5) in \citet{O'Sullivan}, using a half opening angle $2^{\circ}$, a Doppler factor equal to 1.35, a viewing angle $14^{\circ}$, and $\Omega=0.034\,\rm pc\cdot GHz$. The magnetic field strength at the core region is obtained with the relation of $B_{\rm core}=B_1(r/1 {\rm pc})^{-1}$. The magnetic field strength is about 390\,mG for the 8.8\,GHz core at a de-projected distance of $\sim$ 0.02\,pc to the jet apex. It is one orders of magnitude lower than that at a previous event of discrete knot ejection in 2011 \citep{king}, where the magnetic field was estimated from the delay between X-ray flare and radio re-brightening ($1.6 \sim 9.2\,G$), as well as the result ($< 10.2\,G$) from the turnover frequency ($\nu_{\rm peak}\sim 10.4$\,GHz) and the turnover flux $\sim 0.22$\,Jy.

\subsection{Proper Motion} \label{subsec:proper motion}

The outward bright knot has been detected for several months, whose distance to the core is listed in Table \ref{tab:knot}. Although the knot from the initial epoch does not exhibit a discrete Gaussian component, the brightness temperature (see Table \ref{tab:knot}) of the innermost component is higher than that of the other two epochs. This satisfies the characteristic of a cooling component with a decaying brightness temperature. We consider this component to be the same as the discrete knot found in other images.

\begin{figure}
	\centering
	\includegraphics[scale=0.48]{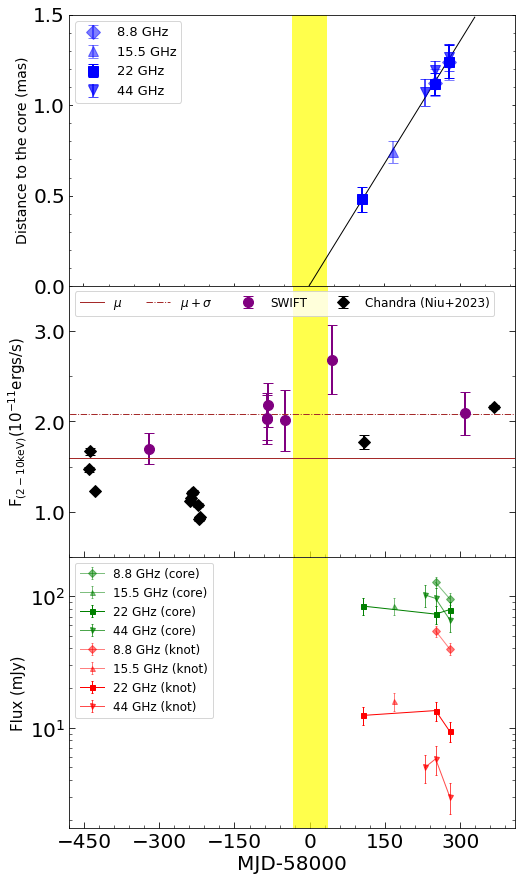}
	\caption{Upper panel: the distance of the discrete knot to the core of 22\,GHz and the linear fitting speed (black line). Middle panel: the X-ray light curve in band 2-10\,keV observed by $SWIFT$ and $Chandra$ \citep{Niu}. The brown solid line and dotted line represent the average flux $\mu$ and the standard deviation $\sigma$, respectively. Bottom panel: the light curve of the core (green) and the knot (red). The yellow region represents the possible date on which the knot passed through the core of 22\,GHz with a confidence of 95\%. \label{fig:motion}}
\end{figure}

We utilize observations at 8.8 and 44\,GHz to enrich the data points. We take into account the core-shift effect for these data, since its magnitude is comparable to the displacement due to proper motion. 
Hence, for each frequency, the knot position is referenced (i.e., shifted) to the location of the core at 22\,GHz using our fitted model with $\Omega$ and $k_{\rm r}$. The distance converted to the core of 22\,GHz is calculated as follows: $d(\nu_2) = d(\nu_1)-\Omega({\nu_2}^{-1/k_{\rm r}}-{\nu_1}^{-1/k_{\rm r}})$, where $\nu_1$ is the observational frequency and $\nu_2 = 22$ (GHz). Figure \ref{fig:motion} upper panel presents the kinematics of the knot following the unified core distance. The linear fit to all data yields an apparent speed of 1.79\,mas/yr (equivalent to $\beta_{\rm app} = 0.1$), significantly slower than $0.51\pm0.17\,c$ previously detected by \citet{king}. There is no indication of acceleration or deceleration in the knot. Under the assumption that the knot maintains a constant intrinsic speed, we deduce the date to pass through the 22\,GHz core to be $\rm MJD\,58020\pm35$.
As Figure \ref{fig:motion} shows, the X-ray light curve shows a flare in MJD\,58045, with a flux approximately twice as high as in the quiescent state reported by \citet{Tomar}. This flux is two standard deviations ($2\sigma$) higher than the average value ($\mu$) in Figure \ref{fig:motion}, while $\mu+\sigma$ is used as a reference for determining whether it is a flare in \citet{Tomar}.
The coincidence of the X-ray flare date and the ejection of the discrete knot suggests a potential correlation between the two events.


The normal of the disk is inclined at an angle of $14\pm 2^{\circ}$ to the line of sight from the spectroscopic observations \citep{devereux}. Here we estimate the viewing angle according to the flux ratio of jet to counterjet.
Even though the counterjet was not detected in all epochs, the flux ratio of the bright knot to the background can be employed to calculate the upper limit of viewing angle. For a discrete knot, the ratio of jet to counterjet is expressed as $R = (\frac{1+\beta \cos \theta}{1-\beta \cos \theta})^{3-\alpha}$, where $R > 13$ is estimated from the image at 22\,GHz with a knot fully resolved from the core. Referring to Figure \ref{fig:sp}, we adopt a spectral index of $-1$ for the knot. This results in an upper limit of the viewing angle at $15^{\circ}$, in accordance with the value in \citet{devereux}. With this angle, its corresponding true speed exceeds $ 0.28\,c$. Another work to explain the jet precession requires a viewing angle greater than $10^{\circ}$ (in preparation by Wu Jiang), equivalent to an upper limit of the true speed $\sim 0.37\,c$. On the other hand, to change the apparent speed from 0.51\,c \citep{king} to 0.1\,c, the viewing angle must be from $14^{\circ}$ to $3^{\circ}$. Under this angle, M81* will present the blazar-like characteristic that the jet has a very strong Doppler-boosting effect and rapid variation.
Given these analyses, we rule out the possibility that the variable apparent speeds observed in 2011 and 2018 can be attributed to different viewing angles. 

\subsection{Polarization} \label{subsec:pol}

\begin{figure}
	\centering
	\includegraphics[scale=0.45]{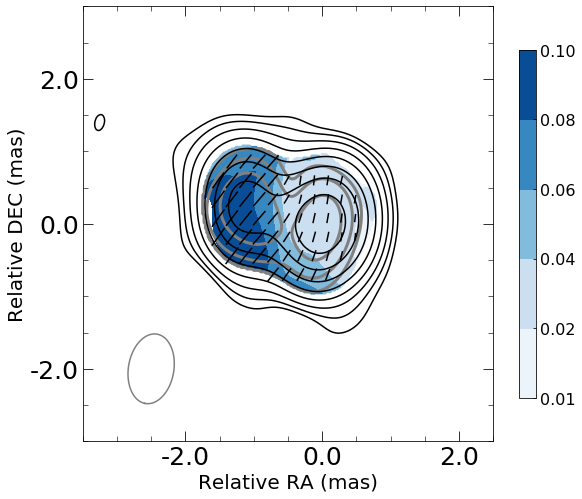}
	\caption{Figure shows the degree of linear polarization with blue colorbar, EVPA and polarization flux with gray contours at 8.8\,GHz observed on June 10, 2018. 
 \label{fig:m81p}  }
\end{figure}

Figure \ref{fig:m81p} shows the jet polarization in the total intensity image at 8.8\,GHz on June 10, 2018. The averaged polarization degree in the core region is $\sim 3\%$, while it is $\sim 8\%$ in the knot region. The electric vector position angle (EVPA) is almost perpendicular to the direction of the jet, suggesting that the poloidal magnetic field dominates this region. It is consistent with the statistics that the spatial distribution of EVPAs in radio galaxies shows an excessive orthogonal polarization orientation based on the linear polarization properties of 436 active galactic nuclei \citep{Pushkarev}. However, the toroidal magnetic field with an average EVPA of $-40^{\circ}$ is comparable to the poloidal field in the knot region. In numerical analysis, continuous jet polarization shows alternating parallel and perpendicular polarization structures \citep{Lyutikov}. The jet with a discrete knot shows differentiated magnetic field compared to continuous one.

\section{Discussions about the possible origins of the discrete knot } \label{sec:discussion}
The distinction of the mechanism of forming discrete knots through observations remains a formidable challenge. 
The unusual low-speed knot in M81* provides new insights into its intrinsic nature. Several interpretations exist for the origin of the discrete knot. We propose three possible origins of the knot to explain our observations. They are a feature of the episodic jet, a low-speed shock wave, and a secondary black hole in a binary system, respectively.

\subsection{Episodic Jet} \label{subsec:model1}

A potential explanation for flares involves episodic jets that are ejected by magnetic forces during magnetic re-connection in accretion flows \citep{Yuan,Cemelji}.  The intrinsic ejection speed of plasmoid can reach 0.08-0.3\,$c$ when launched from the hot accretion flow around the LLAGN M81*.
The adiabatic expansion of the ejected plasmoid undergoes a brightening stage due to the injection of accelerated energetic electrons by magnetic re-connection, which may correspond to the first epoch in our data. During the adiabatic expansion of the ejected plasmoid, the radio flares gradually become optically thin from an initially optically thick emission. The optically thin flares could show strong linear polarization due to the association of the ordered magnetic field structure. An important feature in this scenario is that plasmoid ejection occurs quasi-periodically, with a short period related to the local dynamical orbital timescale at the launching point at about $\sim10 r_{\rm g}$ \citep{Cemelji} from the center black hole, where $r_{\rm g}$ is the gravitational radius \citep{Cemelji}. More importantly, as explored for near-infrared and X-ray flares in Sgr\, A*  \citep{Li2017, Lin}, this model predicates flares duration determined by the local Alfvén timescale, which should be as short as a few hours. However, the period of the X-ray flare and the knot ejection in M81* may be in years according to current observations. 
The scenario of episodic jet associated with magnetic reconnection should be modified to allow for longer period and timescale of flares.

\subsection{Shock} \label{subsec:model2}

Another alternative explanation could be the compression of the magnetic field transverse to the jet caused by a shock. It's possible that the reason why we do not observe the shock front structure in Figure \ref{fig:m81p} is the insufficient angular resolution. 3D MHD simulations show that strong shocks can be produced by the interaction of the jet material with the ambient medium and with the previously injected gas \citep{Dubey}. A precessing jet is less extended and more turbulent. Such a case permits for a non-orthogonal magnetic field to be attributed to the highly asymmetric jet and low relativistic speed \citep{Lyutikov}. In this scenario, a shock wave is generated near the jet base and propagates along the jet direction. The X-ray and radio flares occurred when the shock passed through the high-energy region and the radio core. The time delay as \citet{king} reported between two bands depends on the locations of the high-energy region and the speed of the shock wave. 
Thereafter, a compact feature is observed when the shock wave propagates downstream of the core. The relativistic shock accelerates the electrons in the jet and amplifies the magnetic field. Simulation shows that shocks in the jet with a precessing nozzle accelerate electrons more efficiently than with a steady nozzle or a time-variable nozzle \citep{Dubey}. For electrons with a power law distribution, only those with $\gamma_e \geq \Gamma_s m_p/m_e$ can be accelerated by shock \citep{Lyutikov}, where $\Gamma_s$ is the shock Lorentz factor, $m_p$ and $m_e$ are the proton and electron masses.
It means that a slower shock wave can accelerate more relativistic electrons. This explains why the knot with a lower speed has a higher flux compared with the observations in 2011. In addition, the magnetic field estimated during the flare \citep{king} was amplified by the shock, while the field strength estimated from the core shift in this work corresponds to the centimeter VLBI core region at the quiescent state. 

\subsection{Binary Black Hole} \label{subsec:model3}

The discrete knot could be associated with the ongoing circumbinary accretion around the massive black-hole binary. For the equal-mass ratio massive binary in eccentric orbit, accretion onto the primary and secondary component of the binary will experience a symmetry breaking, with the secondary black hole accreting about one order of magnitude than the other, and this accretion pattern is reversed back on the long-term precession timescale \citep[e.g., see review by][]{Gold,Lai2022}. For unequal-mass eccentric binaries, it is also possible that the secondary massive black hole accretes much less mass than the primary when the circumbinary disk undergoes forced precession, which happens for an even larger orbital eccentricity \citep{Gold1,Siwek2023}.
The significant suppression of accretion rates (or supply rate) onto the less massive member of the binary results in the synchrotron self-absorption frequency shifting toward much lower frequency. It is thus very likely that the core of the more massive black hole appears as optically thick while the dimmer knot associated with the less massive black hole is instead optically thin. This spectral index asymmetry is consistent with the observations shown in Figure~\ref{fig:sp}.

Interestingly, the recent discovery of periodic radio and X-ray outbursts, as well as jet wobbling associated with the core show the observational evidence for this massive binary black hole scenario (Jiang et al. 2023, in preparation). The binary semi-major axis is constrained to be $0.02$ pc (or equivalently $\sim 1$ mas at the distance of M81), which is in good agreement with the maximum distance between the core and the knot as shown in Figure~\ref{fig:motion}. 
The almost linear projected proper motion of ejected knots up to $\sim 1\ {\rm mas}$ as shown in Figure~\ref{fig:motion} further suggests a large orbital eccentricity for the binary. This is consistent with the small mass ratio of $q\sim0.1$ constrained from the periodic outburst observations, which requires a highly eccentric binary to suppress the accretion onto the secondary as mentioned above. 
The highly eccentric binary can also induce significant orbital velocity variation at different orbital phases. 
However, a robust circumbinary accretion disk theory with realistic parameter space is still lacking, which prevents us predicting quantitatively the accretion rates onto each component of the binary. 
More importantly, current sparse time-domain sampling observations suffers from the model degeneracy between the ballistic motion of ejected plasmoid and highly eccentric orbital motion of the binary.  
The polarization signature from the companion depends sensitively on the magnetic field geometry of the emission region, which is largely uncertain. We could expect different field geometries between the core and the knot since their emission could be dominated by different regions of the accretion-jet system \citep{Farris,Gold1}.


\section{Summary} \label{sec:summary}

In this paper, we present multifrequency VLBI and contemporary X-ray observations of one of the closest LLAGN M81*. A new discrete knot was detected at multi-epochs. Our results are summarized as follows:

Taking advantage of the optically thin knot, we fit the core shift relation of $r_{\rm core}=1.72\nu^{-1/1.12}$ up to 44\,GHz. This relation suggests that the position of the central engine is located at $\sim0.06$\,mas upstream of the core of 44\,GHz. This result is well consistent with what has been reported in \citet{mart}. The magnetic field strength at 8.8\,GHz VLBI core is hundreds of milliGauss.

The distances from the knot to the core of 22\,GHz increase with respect to time, corresponding to a proper motion of $\sim$1.79\,mas/yr (0.1\,$c$), which is much lower than the speed of another discrete knot observed in 2011 (0.51\,$c$) \citep{king}. We can exclude the change of viewing angle ($10^{\circ}$-$15^{\circ}$) in affecting the apparent speed and estimate a true speed of 0.28-0.37\,$c$.

From the proper motion of the knot, we retraced the time when the knot passed through the radio core. This time (MJD\,$58020\pm35$) is associated with a moderate X-ray flare in MJD\,58045. The connection of the X-ray flare and the ejection of the discrete knot helps to reveal the nature of the knot. We propose three possible origins of the knot to explain our observations: an episodic jet ejection, a shock wave, and a secondary black hole in a binary system.

In the scenario of the episodic jet ejection, plasmoid in accretion flows is ejected by magnetic forces during magnetic re-connection to form the discrete knot \citep{Yuan,Cemelji}. An event of magnetic re-connection is accompanied by the X-ray flare. In the scenario of shock wave, the X-ray flare occurred when the shock wave passed through the high-energy region at upstream of the radio core. It is similar to the connection between high-energy flares and the emergence of new components in \citet{Rani}. The low-speed knot is possibly associated with the secondary of the massive black hole binary with a small mass ratio of $q\sim 0.1$ and a large orbital eccentricity. It should be noted that these three scenarios could occur simultaneously as a binary sweeping through a disk could also provide jetted outflows and shocks.

It's still a challenge to distinguish the origins of the knot from current data. Some signatures can be used to distinguish these origins. The opacity for episodic jet evolves with its volume expansion, which leads to a much shorter time delay between flares in the X-ray and the optical/infrared for episodic jets than that for the shock model. Additionally, the time duration of the flare may be different for these two models. The high energy electrons in episodic jets will cool down more quickly, therefore, its flare timescale should be shorter. We can identify the binary black holes by detecting a periodic jet wobbling, as well as the possible time-variant positions of the core through high-precision VLBI phase-referencing observations, if the black holes wobble around the mass center of the binary system. Further multifrequency VLBI imaging observations with full-polarizations are thus required to monitor the proper motion of possible knots, the radio spectra, and polarization signatures. Especially those data at the flare stage are important to distinguish different scenarios.

\begin{acknowledgments}
The authors thank the anonymous referee for the helpful suggestions. This work was supported in part by the National Natural Science Foundation of China (Grant Nos. 12173074, 11803071, 12192223), and the Natural Science Foundation of Shanghai (Grant NO. 23ZR1473700). The authors thank the observation facilities of VLBA, SWIFT and all people working for observation and correlation. The authors thank the VLBA data archive maintained by NRAO for the essential data. VLBA and NRAO are facilities of the National Science Foundation operated under a cooperative agreement by Associated Universities, Inc. We acknowledge the MOJAVE database for querying the data, which is maintained by the MOJAVE team. This study uses data from the VLBA-BU Blazar Monitoring Program, funded by NASA through the Fermi Guest Investigator Program. 
IMV acknowledges partial support from Generalitat Valenciana (GenT Project CIDEGENT/2018/021), the MICINN Research Project PID2019-108995GB-C22 and the ASTROVIVES FEDER infrastructure IDIFEDER-2021-086.

\end{acknowledgments}

\bibliography{sample631}{}
\bibliographystyle{aasjournal}

\clearpage

\end{document}